\documentclass{jpsj-suppl}
\usepackage{txfonts}
\usepackage{times}
\usepackage{color}
\newcommand{\ltsim}{\protect\raisebox{-0.5ex}{$\:\stackrel{\textstyle <}{\sim}\:$}}

\title{Dynamical mean-field analysis of ordered phases \\
in the half-filled Holstein-Hubbard model}

\author{Yuta Murakami, Philipp Werner$^*$, Naoto Tsuji, Hideo Aoki}
\inst{Department of Physics, University of Tokyo, Hongo, Tokyo 113-0033, Japan, \\
$^*$
Department of Physics, University of Fribourg, 1700 Fribourg, Switzerland}

\abst{
We investigate ordered phases (s-wave superconducting (SC),  antiferromagnetic (AF) and  charge ordered (CO) phases) of the half-filled Holstein-Hubbard model using the dynamical mean-field theory in combination with a continuous time quantum Monte Carlo impurity solver.
We particularly focus on the parameter regime where the electron-electron Coulomb interaction and the phonon mediated retarded attractive interaction 
are both comparable to the bare electron bandwidth to understand the interplay of these interactions and the retardation effects due to
finite phonon frequency $\omega_0$.  
Results for the transition temperatures for SC and AF as a function of the effective interaction
$U_{\rm{eff}}$ ($=U-\lambda$ with $U$ the Coulomb repulsion and $\lambda$ the phonon-mediated attraction) show deviations from the the corresponding results in the 
anti-adiabatic limit.
These are analyzed by comparing with an effective static model in the polaron picture. We also determine the phase diagram around $U_{\rm{eff}}=0$ at low, but nonzero temperatures
to show that paramagnetic phases appear between AF and CO when $U$ and $\lambda$ are small or large, while AF and CO 
directly compete with each other at moderate $U$ and $\lambda$.}

\kword{Holstein-Hubbard model, dynamical mean-field theory, 
superconductivity, antiferromagnetism, charge order}

\begin{document}
\maketitle

\section{Introduction}
The interplay of electron-electron Coulomb interactions and electron-phonon interactions brings about rich physics.  
Interesting examples are carbon-based superconductors such as alkali-doped fullerides, which accommodate s-wave superconductivity (SC). 
They exhibit a $T_c$ dome in the temperature-pressure phase diagram with 
an antiferromagnetic (AF) phase adjacent to SC \cite{c60_1,c60_2,c60_3}. 
It is believed that in these compounds the electron-electron Coulomb interaction, the electron-phonon interaction and the phonon frequency are all comparable 
 to the electron bare bandwidth. Other examples are the recently discovered aromatic superconductors, which also belong to a similar parameter regime \cite{picenePCCP,piceneKosugi,picene1,picene2,picene3}.

To give physical insights to this class of compounds, we consider  
the Holstein-Hubbard (HH) model, which is the simplest model that 
has both a local electron-electron Coulomb interaction and a local electron-phonon interaction. The Hamiltonian is 
\begin{align}
 H=&-t\sum_{\langle i, j\rangle,\sigma}(c^{\dagger}_{i\sigma}c_{j\sigma}+\text{H.c.}) + \sum_i[Un_{i\uparrow}n_{i\downarrow}-\mu(n_{i\uparrow}+n_{i\downarrow})]\nonumber\\
 &+g\sum_i(b^{\dagger}_i+b_i)(n_{i\uparrow}+n_{i\downarrow}-1)+\omega_0\sum_ib^{\dagger}_ib_i,
  \label{eq:HHmodel}
\end{align}
where $c^{\dagger}_{i\sigma}$ creates an electron at the $i$th site with spin $\sigma$, $t$ is the hopping parameter, $n_{i\sigma}=c^{\dagger}_{i\sigma}c_{i\sigma}$, $U$ is the on-site Coulomb interaction, 
$b^{\dagger}$ creates a phonon, $g$ is the electron-phonon coupling constant, 
and $\omega_0$ is the frequency of a (dispersionless) phonon.

When we integrate out the phonon degrees of freedom in a path-integral framework, we obtain an effective retarded electron-electron interaction 
\begin{equation}
 U_{\mathrm{eff}}(\omega) = U-\frac{2g^2\omega_0}{\omega_0^2-\omega^2}.
\end{equation}
In the low-energy regime we have
\begin{equation}
U_{\mathrm{eff}}\equiv  U_{\mathrm{eff}}(\omega=0) \equiv U-\lambda, \quad
\lambda=2g^2/\omega_0,
\end{equation} 
which can be regarded as a measure of the effective electron-electron interaction. In the anti-adiabatic limit of $\omega_0\rightarrow \infty$ with $\lambda$ and $U$ fixed, the HH model 
is reduced to the Hubbard model with the static interaction $U_{\rm{eff}}$. This anti-adiabatic description is sometimes considered as approximate treatments for finite $\omega_0$ \cite{c60_2}, 
and it is important to examine the deviations from this simple anti-adiabatic description \cite{HH2d6}.

The Holstein-Hubbard model has been investigated in various spatial dimensions \cite{1dimdmrg1,HH2d6,HHdmft4,HHnrg,HHdmft6,HHnrg2,HHdmft1,HHdmft2,HHdmft3,HHdmft7}. However, the nature of the ordered phases is not fully understood  so far, 
specifically in the parameter regime where $U$ and $\lambda$ are comparable to the bandwidth. The questions we want to address in this paper are: (i) what is the effect of the coexistence of two types of interactions and of the retardation arising from a finite value of $\omega_0$, (ii) to what extent the anti-adiabatic description is valid when $\omega_0$ is finite, (iii) whether we can find a better approximate description, and (iv) how the ordered phases compete with each other. In order to answer these questions, we study the half-filled Holstein-Hubbard model in infinite spatial dimensions. We compute the transition temperatures for SC and AF, which are analyzed with an effective static model in the polaron picture introduced in Sec. 2. In addition, we map out the phase diagram around $U_{\rm{eff}}=0$ at nonzero temperatures.
Detailed discussions of these topics have been published in Ref.~\cite{Murakami2013}. Here we add a discussion for the isotope effect on the SC phase, a systematic analysis of the AF phase and a more elaborate study of the behavior at higher temperatures around $U_{\rm{eff}}=0$.

\section{Method}
\subsection{Dynamical mean-field theory}
In order to solve the model,
we employ the dynamical mean-field theory (DMFT), which is exact in infinite spatial dimensions, with the continuous-time quantum Monte Carlo method (CT-QMC) 
as an impurity solver \cite{Werner2006, HHdmft4}. 
In the case of infinite spatial dimensions, the problem becomes a self-consistent impurity problem \cite{dmft4}.
In our case, the effective impurity problem reads
$H_{\rm{imp}}=H_{\mathrm{loc}}+H_{\mathrm{bath}}+H_{\mathrm{mix}}$ 
with 
  \begin{align}
   H_{\mathrm{loc}}=&Un_{\uparrow}n_{\downarrow}
-\mu(n_{\uparrow}+n_{\downarrow})+g(b^{\dagger}+b)(n_{\uparrow}+n_{\downarrow}-1)
   +\omega_0b^{\dagger}b,\label{H_imp1}\\
     H_{\mathrm{bath}}=&\sum_{p,\sigma}\epsilon_p c^{\dagger}_{p,\sigma}c_{p,\sigma}+\sum_{p}(\Delta_p c^{\dagger}_{p\uparrow}c^{\dagger}_{-p\downarrow}+{\rm H.c.})\label{H_imp2},\\
     H_{\mathrm{mix}}=&\sum_{p,\sigma}(V_p^{\sigma}d^{\dagger}_{\sigma}c_{p,\sigma}+{\rm H.c.}),
     \label{H_imp3}
  \end{align}
    where $H_{\mathrm{loc}}$ denotes the Hamiltonian of the impurity site with $n_\sigma=d_\sigma^\dagger d_\sigma$, $H_{\mathrm{bath}}$ describes a superconducting bath with
  finite $\Delta_p $, and $H_{\mathrm{mix}}$ denotes the hybridization between the bath and impurity.
  The electrons have a local Coulomb interaction at the impurity site and are coupled to local phonons. We use the infinite-coordination Bethe lattice, 
  which is bipartite and has a semicircular density of states, $\rho(\epsilon)=\frac{1}{\pi t}\sqrt{1-[\epsilon/(2t)]^2}$.  
Hereafter we take the quarter of the bandwidth ($t=W/4$) as the unit of energy.  
Based on Ref.~\cite{HHdmft4}, we have developed a CT-QMC (hybridization expansion) solver for this impurity problem \cite{Murakami2013}. When we treat a charge-ordered phase (CO) or an antiferromagnetic phase (AF), we assume commensurate checkerboard order. 

\subsection{Effective static model in a polaron representation}
An effective static model in a polaron picture can be obtained as follows \cite{Casula2012}. First, we perform the Lang-Firsov (LF) canonical transformation, $H_{\rm{LF}}\equiv e^{S} H e^{-S}$ with 
$S=(g/\omega_0) \sum_{i}(n_i-1)(b_i^{\dagger}-b_i)$.  We note that, since this transformation changes the electron creation operator to $e^{S} c^{\dagger} e^{-S}=e^{(g/\omega_0)(b^{\dagger}-b)}c^{\dagger}$, the creation operator $c^\dagger$ after 
the LF transformation has a meaning of creating an electron dressed by phonons (i.e. a polaron).  
The second step is to make a projection onto the subspace of zero phonons 
by assuming that $\omega_0$ is large enough and the temperature low enough that phonon excitations hardly occur, i.e., $H_{\rm{eff}}=\langle 0|H_{\rm{LF}}|0\rangle$ with $|0\rangle$ 
being the phonon vacuum. The resulting Hamiltonian is \cite{Murakami2013}
  \begin{align}
  H_{\mathrm{eff}}=&-Z_{\mathrm{B}}t\sum_{\langle i,j\rangle,\sigma}(c^{\dagger}_{i\sigma}c_{j\sigma}+\text{H.c.}) + U_{\mathrm{eff}}\sum_{i}n_{i\uparrow}n_{i\downarrow}-\mu_{\mathrm{eff}}\sum_i n_i
\label{eq:eff}
  \end{align}
  with $Z_B=\exp(-g^2/\omega_0^2)$ and $\mu_{\rm{eff}}=\mu-g^2/\omega_0$.
  For this effective model, the transition temperatures for ordered phases scale as
  \begin{align}
  T_c[U, U_{\mathrm{eff}}, Z_{\mathrm{B}}]&\approx Z_\mathrm{B} T^0_{c}[U_{\mathrm{eff}}/Z_\mathrm{B}], \label{eq:Teff}
  \end{align}
  where $T^0_{c}[U]$ denotes the transition temperature (for SC if $U<0$ and AF if $U>0$) of the Hubbard model with hopping $t$ and static interaction $U$.
  We can also derive other physical quantities such as  the superconducting order parameter from the effective polaron picture (see Ref.~\cite{Murakami2013} for details).

\section{Results}
In order to determine the phase boundary, we computed the order parameter for different temperatures inside the ordered phase and extrapolated these curves to zero. 
\subsection{Transition temperature for superconductivity}
\begin{figure}[h]  
\begin{center}
\includegraphics[width=15.5cm]{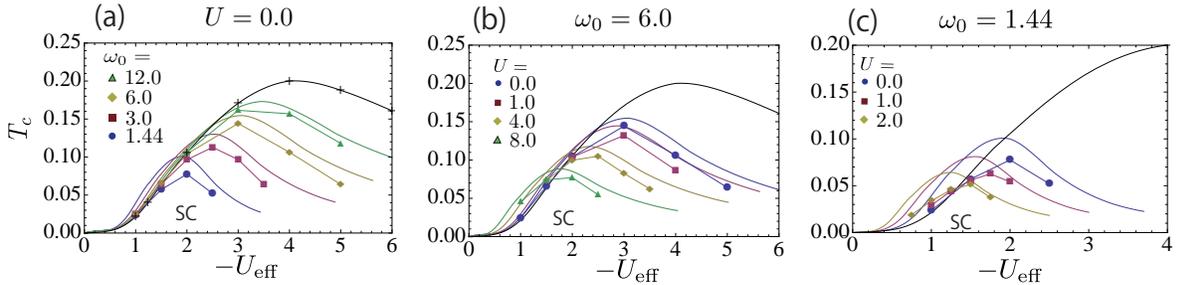}
\end{center}
\caption{Phase diagram for SC against $T$ and $-U_{\rm{eff}}$ for various sets of parameters. Markers are DMFT+QMC results, while curves without markers 
represent the results in the effective polaron model [Eq.~(\ref{eq:eff})].  Crosses in (a) represent the result for the attractive Hubbard model with $U_{\mathrm{eff}}$, and  the black curves in each panel are the interpolation of these data. Note the different ranges in plot (c). These figures are taken from Ref.~\cite{Murakami2013}.}
\label{fig:tc_sc}
\end{figure}
Figure \ref{fig:tc_sc} displays the phase diagram against $-U_{\rm{eff}}$ and $T$ for various sets of parameters. In all cases, $T_c$ increases with $-U_{\rm{eff}}$ in the small $|U_{\rm{eff}}|$ regime and decreases in the large $|U_{\rm{eff}}|$ regime. This gives a $T_c$ dome, which manifests a BCS-BEC crossover. In the strong-coupling regime, bipolarons are created due to the strong effective attraction between electrons.

Figure \ref{fig:tc_sc}(a) illustrates the effect of the retardation coming from finite $\omega_0$ at $U=0$. We find that, as $\omega_0$ decreases, the height of the $T_c$ dome decreases with the peak position shifting to the weak $|U_{\rm{eff}}|$ regime. Thus the deviation from the attractive Hubbard model with $U_{\rm{eff}}$ is significant when $|U_{\rm{eff}}|$ is large.
On the other hand, in the weak $|U_{\rm{eff}}|$ regime, $T_c$ is well described by the attractive Hubbard model, although this regime shrinks as $\omega_0$ decreases.
Panels (b) and (c) show the results obtained when the Coulomb interaction is switched on with fixed $\omega_0=6$ (b) and $\omega_0=1.44$ (c).  For a fixed
value of $\omega_0$, one can observe that the position of the $T_c$ dome again shifts to the weak-coupling regime with its height reduced when $U$ increases. Here we note that in this case $T_c$ in the weak-coupling regime is not necessarily well approximated by the attractive Hubbard model; the critical temperature is higher than the $T_c$ for the Hubbard model if plotted as a function of $|U_{\rm{eff}}|$. 

Our results give insights into the isotope effect with large phonon frequency.
$g$ and $\omega_0$ in the HH model usually scale with a mass ($M$) of the atoms as $g\sim M^{-1/4}$ and $\omega_0\sim M^{-1/2}$, while  $\lambda$ and $U$ are 
expected not to change. Our results suggest that, when $U$ is small, we can expect that 
$T_c$ do not significantly change in the weak-coupling regime (Fig. \ref{fig:tc_sc}(a)). On the other hand, if $U$ is large, $T_c$ {\it increases} in the weak-coupling regime for heavier atoms (Fig. \ref{fig:tc_sc}(b)(c)), which is opposite to the conventional isotope effect with slow phonon frequencies. We can see this by comparing the result for a finite $\omega_0$ and a certain $U_{\rm{eff}}$ and $U$ to the result for the attractive Hubbard model with the same $U_{\rm{eff}}$ (see for example the data for $U_{\rm{eff}}=1,U=2,\omega_0=1.44$ in Fig. \ref{fig:tc_sc}(c)).

The effective polaron model [Eq.~(\ref{eq:eff})] describes the effect of the retardation and the Coulomb interaction rather well (Fig.~1).  
It tends to overestimate $T_c$ in the parameter range investigated.  
The effective model gives more accurate results for larger $\omega_0$. At $\omega_0=6$ and $2\ltsim U_{\rm{eff}}/Z_B\ltsim 6$, we find a minimum of $\delta T_c\equiv |T_c-T_{c,\text{eff}}|/T_c\sim0.1$ around $U_{\rm{eff}}/Z_B\simeq4$, 
at least up to $U=8$.  
For $\omega_0\ge4$, $\delta T_c$ shows a relatively weak dependence on $U_{\rm{eff}}$.  We note that when $\omega_0\ge4$ the effective model [Eq.~(\ref{eq:eff})] is quantitatively good in a wide range of parameters, and it qualitatively captures the effect of the retardation and the Coulomb interaction even at $\omega_0$ as 
small as $\sim1.44t$.

\subsection{Transition temperature for antiferromagnetism}
\begin{figure}[t]  
\begin{center}
\includegraphics[width=15.5cm]{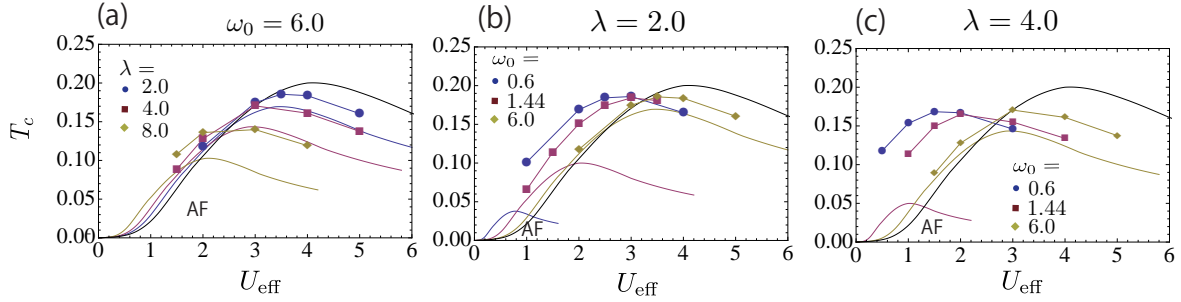}
\end{center}
\caption{Phase diagram for AF against $T$ and $U_{\rm{eff}}$ for various sets of parameters. Markers are DMFT+QMC results, while curves without markers represent the results for the effective polaron model [Eq.~(\ref{eq:eff})]. The  black curve in each panel 
represents the transition temperature in the attractive Hubbard model with $U_{\mathrm{eff}}$. }
\label{fig:tc_af}
\end{figure}

Now we turn to the AF phase, which has not been discussed systematically in Ref.~\cite{Murakami2013}.
Since AF originates from the repulsive 
interaction, a question of interest here concerns the effect of the retarded attractive interaction mediated by the phonons. Figure \ref{fig:tc_af} (a) shows the effect of the strength of the attractive interaction ($\lambda$) with $\omega_0$ fixed. As $\lambda$ becomes large the AF dome shifts to the small $U_{\rm{eff}}$ regime and the height decreases. When we change $\omega_0$ 
with $\lambda$ fixed to extract 
the effect of the retardation in Fig. \ref{fig:tc_af} (b)(c), we find that the peak shifts to the small $U_{\rm{eff}}$ regime, while the height of the peak shows a small dependence on $\omega_0$ for $0.6 \leq \omega_0 \leq 6.0$.

The effective model [Eq.~(\ref{eq:eff})] also predicts a shift of the peak to the small $U_{\rm{eff}}$ regime for smaller $\omega_0$ or for larger $\lambda$ (curves without markers in Fig. \ref{fig:tc_af}). It turns out that the effective model always underestimates the AF transition temperature, and that  the height of the dome is less sensitive to changes in $\omega_0$ or $\lambda$ than what the effective model predicts. 
For $2\ltsim U_{\rm{eff}}/Z_B\ltsim6$ at $\omega_0=6$ and $\omega_0=1.44$, it turns out  that the accuracy of the effective model gets worse as $U_{\rm{eff}}/Z_B$ becomes larger, $\lambda$ larger, or $\omega_0$ smaller. 
At $\omega_0=6, \lambda=4$ for example, the deviation ($\delta T_c$ ) in the AF 
transition temperature monotonically increases from $\delta T_c\sim0.1$ to $0.25$.

\subsection{Phase diagram around $U_{\rm{eff}}=0$}
\begin{figure}[t]  
\begin{center}
\includegraphics[width=15.5cm]{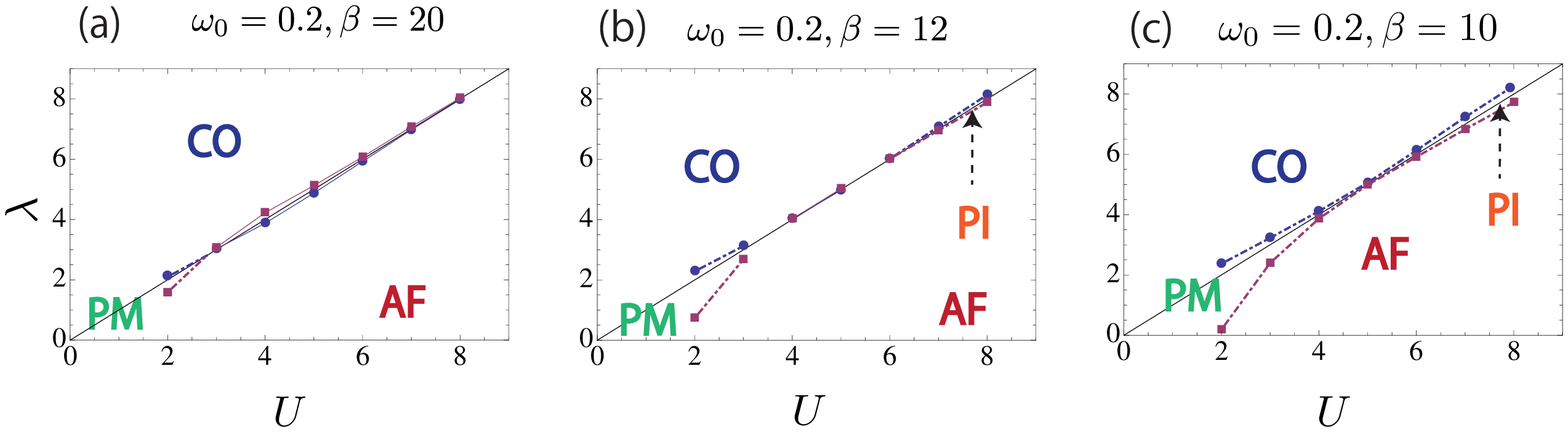}
\end{center}
\caption{Phase diagram at finite temperatures in the $U-\lambda$ plane.  Dotted lines indicate boundaries between an ordered phase and a paramagnetic phase with a continuous transition. Solid lines represent the boundaries of the region where stable solutions of CO or AF exist. Blue is for CO and red is for AF. Panel (a) and (b) are quoted from Ref.~\cite{Murakami2013}.}
\label{fig:af_co}
\end{figure}

\begin{figure}[t]  
\begin{center}
\includegraphics[width=13cm]{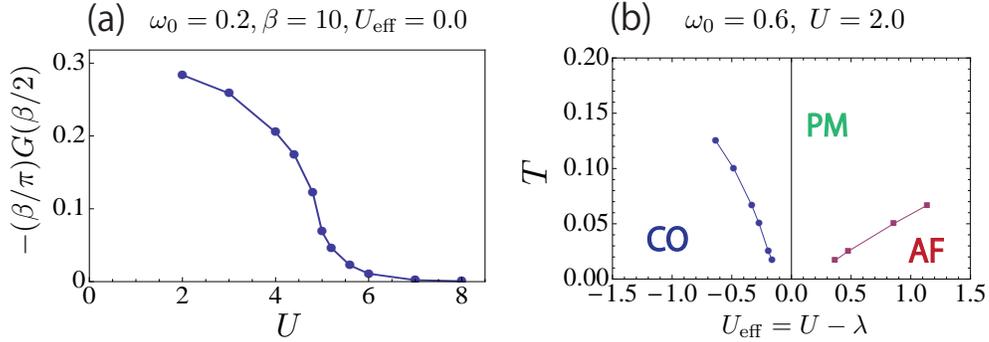}
\end{center}
\caption{(a) $-(\beta/\pi)G(\tau=\beta/2)$ plotted against $U$ for 
$U_{\rm{eff}}=0$ at $\omega_0=0.2,\beta=10$. (b) Phase diagram at finite temperatures in the $T-U_{\rm{eff}}$ plane at $U=2,\omega_0=0.6$. Panel (b) is taken from Ref.~\cite{Murakami2013}.}
\label{fig:fig4}
\end{figure}

We have obtained the phase diagram around $U_{\rm{eff}}=0$ allowing for AF, CO and SC order.  Results for $\omega_0=0.2$ and different low, but nonzero temperatures  are shown in Fig.~\ref{fig:af_co}. We have found no SC phase, 
while there is a paramagnetic metallic phase (PM) between CO and AF (panels (a), (b) and (c)) when $\lambda$ and $U$ are small enough. When both parameters are large, 
on the other hand, we find a paramagnetic insulating phase (PI) around $U_{\rm{eff}}=0$ (panels (b) and (c)). The transitions between the ordered phases and the paramagnetic phases are continuous. In the intermediate-coupling regime and at low enough temperatures, the transition between AF and CO is direct and of first order. In other words,
there is a region where both AF and CO are stable solutions of the DMFT equations. This region is always around $U_{\rm{eff}}=0$ (as pointed out in Ref.~\cite{HHdmft3} for $T=0$), even at nonzero temperatures. The paramagnetic phases become wider as $T$ increases, while the hysteretic region shrinks. 
Eventually, at higher $T$, the  PM and PI are connected around $U_{\rm{eff}}=0$ (panel (c)).
In order to obtain information on the density of state (DOS) at the Fermi energy, we can use a relation between the Matsubara Green's function and the spectral function $A(\omega)$,
$G(\tau=\beta/2)=-\int  d\omega \frac{1}{2\cosh(\beta\omega/2)}A(\omega)$.
This formula implies that if the temperature is low enough, $-(\beta/\pi)G(\tau=\beta/2)$ becomes a good approximation for $A(\omega)$. The result in the paramagnetic regime ($U_{\rm{eff}}=0$)
is shown for different values of $U$ and  $\omega_0=0.2$, $\beta=10$ in Fig. \ref{fig:fig4} (a).  
As can be seen, the paramagnetic phase has a nonzero DOS at the Fermi energy for small $U$, which indicates 
a metallic behavior. On the other hand, for larger $U$ the DOS rapidly decreases to almost zero, which indicates an insulating behavior. Our analysis shows that a metal-Mott insulator crossover occurs within the paramagnetic region of Fig.~\ref{fig:af_co}(c).

Finally, in Fig.~\ref{fig:fig4}(b), we show the typical temperature dependence of the PM solution in the weak-coupling regime. Here we fix $\omega_0=0.6$. Even though we cannot go 
to lower temperatures, the data seem to suggest that there is no hysteretic region for AF and CO, and that the qualitative nature of the phase transition between AF and CO at $T=0$ is different in the weak- and intermediate-coupling regimes, as suggested in Ref.~\cite{HHdmft3}.

\section{Conclusion}
We have investigated the ordered phases in the half-filled Holstein-Hubbard model, using the dynamical mean-field theory, which is reliable in high dimensions, with a continuous time quantum Monte Carlo impurity solver.
In particular, we have revealed the behavior of the transition temperatures of SC and AF when $\lambda, U$ and $\omega_0$ are comparable to the electronic bandwidth  $W$.  We have shown deviations from the anti-adiabatic description (the Hubbard model) at finite $\omega_0$, and discussed the accuracy of the effective polaron description for each phase. We have also obtained the phase diagram around $U_{\rm{eff}}=0$ at nonzero temperatures, showing
that the paramagnetic metallic (insulating) phase appears when $U$ and $\lambda$ are small (large) enough. 

The authors would like to thank T. Kariyado, T. Oka, M. Tezuka and A. Koga for 
discussions.  H.A. and N.T. have been supported by
LEMSUPER (EU-Japan Superconductor Project) from JST. PW acknowledges support from SNF Grant 200021-140648.

\end{document}